\input{aipcheck}

\documentclass[
     ,final            
  ]
  {aipproc}

\layoutstyle{6x9}
\usepackage{amsmath}
\usepackage{amssymb}

\begin{document}

\title{Extraction of neutron structure from tagged structure functions}


\author{W. Cosyn}{
  address={Department of Physics, Florida International University, Miami, Florida 33199, USA},
  altaddress={On leave from: Department of Physics and Astronomy,
 Ghent University, Proeftuinstraat 86, B-9000 Gent, Belgium}
}

\author{M. Sargsian}{
  address={Department of Physics, Florida International University, Miami, Florida 33199, USA}
}

\begin{abstract}
We present work in a model used to describe semi-inclusive deep inelastic
scattering off the deuteron.  The model uses the virtual nucleon approximation
to describe the interaction of the photon with the bound neutron and the
generalized eikonal approximation is applied to calculate the final-state
interaction diagram.  Comparison with data taken at Jefferson Lab shows good
agreement in the covered range of kinematics and points at a largely
suppressed off-shell rescattering amplitude.  The 
$W$ and $Q^2$ dependences 
of  the total cross section and slope factor of the interaction of 
DIS products, $X$,   off the spectator nucleon are extracted.  Starting from
the JLab data and our model calculations, we outline and apply an extrapolation
method to obtain the neutron structure function $F_{2N}$ at high Bjorken $x$.
\end{abstract}

\maketitle


\section{Introduction}

A reaction that can be used to study the influence of QCD dynamics  at
nucleonic length scales is semi-inclusive deep inelastic scattering off the
deuteron ($d(e,e'p_s)X$), where a spectator proton is detected in the
final state. At very low spectator momenta the neutron is nearly on-shell and
the reaction can be used to extract information about the ``\emph{free}''
neutron structure function in a way that minimizes the nuclear binding effects
inherent in scattering off a deuteron target.  At higher spectator momenta the
reaction can be used to study the modifications of nucleon properties and the
role quark degrees of freedom in high density configurations of the deuteron. 
In kinematics wich favour final-state interactions (FSI), the space-time
evolution of hadronization can be examined.  Two
recent Jefferson Lab Hall B experiments have studied the reaction: one at high
\cite{Klimenko:2005zz}, and the other  at low spectator momenta
\cite{Bonus:2003}.

In order to provide meaningful interpretations of the measured data,
theoretical models that quantify the importance of the FSI in the reaction are
needed.  The major difficulty in doing this is that one lacks detailed
information about the composition and space-time evolution of the hadronic
system produced in the deep inelastic scattering and how this changes as a
function of Bjorken $x$ and $Q^2$.  Several theoretical approaches that provide
models for the  $d(e,e'p_s)X$ reaction can be found in the literature
\cite{Simula:1996xk,Melnitchouk:1996vp,Sargsian:2005rm,CiofidegliAtti:1999kp,
CiofidegliAtti:2002as,CiofidegliAtti:2003pb,Palli:2009it,Atti:2010yf}.  In the
following, we will present work in a model based on the general properties of
soft rescattering \cite{Cosyn:2010ux}.  Results are compared to data taken in
the JLab Deeps experiment and we use these results and data to describe a
method to extract the neutron structure at high $x$.

\section{Formalism}

The interaction of the virtual photon with the bound nucleon is treated with
the \emph{virtual nucleon approximation} (VNA)
\cite{Melnitchouk:1996vp,Sargsian:2005rm,Sargsian:2001gu}.  In the VNA, the
spectator nucleon is taken on the mass shell, while the virtual photon interacts
with an off-shell nucleon.  The VNA is based on the following main assumptions:
(i) only the $pn$ component
of the
deuteron wave function is considered in the reaction, 
(ii) the negative energy projection of the virtual
nucleon propagator gives negligible contribution to the scattering amplitude,
and (iii) interactions of the virtual photon with exchanged mesons is 
neglected. Assumptions (i) and (ii) can be satisfied when the momentum of the
spectator proton is limited to $p_s \leq 700$ MeV/c \cite{Sargsian:2009hf}, 
while (iii) is satisfied at large $Q^2$ ($> 1$ GeV$^2$)
\cite{Sargsian:2001ax,Sargsian:2002wc}.

The nuclear  wave  function in the VNA is  normalized to account for the baryon
number 
conservation \cite{Frankfurt1981215,Frankfurt:1976gz,Frankfurt1987254,
Landshoff:1977pg}:
\begin{equation}
\int \alpha |\Psi_D(p)|^2 d^3p  = 1,
\label{norm}
\end{equation}
where $\alpha= 2-\frac{2(E_s-p_{s,z})}{M_D}$ is the light cone momentum
fraction of the deuteron 
carried by the bound nucleon normalized in such a way that the half of the
deuteron 
momentum fraction corresponds to $\alpha=1$.
Because of the virtuality of interacting nucleon it is impossible 
to satisfy the momentum  sum rule at the same time, which can be qualitatively 
interpreted as part  of the deuteron momentum
fraction being 
distributed to non-nucleonic degrees of freedom.

In the model, two Feynman diagrams are taken into account: the plane-wave
diagram where the on-shell proton is a pure spectator and the FSI diagram
where a soft rescattering occurs between the produced $X$ and the spectator.  To
describe the FSI of the produced $X$ with the spectator nucleon, the
\emph{generalized eikonal approximation} is applied. At the energies under
consideration in the JLab experiments, we can assume the rescattering to be
diffractive and occurring over small angles.  The following eikonal form is
adopted for the $X-N$ rescattering amplitude:
\begin{equation} \label{eq:ampl}
 f_{X^\prime N, X N} = \sigma(Q^2,x_{Bj})(i + \epsilon(Q^2,
x_{Bj}))e^{\frac{B(Q^2,x_{Bj})}{2} t},
\end{equation}
where  $\sigma$, $\epsilon$ and $B$ are the effective total cross section, real
part and the slope factor 
of the diffractive $X^\prime N\rightarrow X N$ scattering amplitude.
In the derivation of the FSI amplitude, a factorized approach is used, whereby
the interaction of the virtual photon with the off-shell nucleon is taken out
of the integration over the intermediate spectator momentum.  Such an
approximation is also applied in quasi-elastic scattering
\cite{Jeschonnek:2000nh,Sargsian:2009hf}, being valid up to spectator momenta
around 400 MeV/c.

We define
the lab frame
four-momenta of the involved particles as $p_D\equiv(M_D,0)$ for the deuteron, $
q\equiv (\nu,\vec{q})$ for the virtual photon (with the z-axis chosen along
$\vec{q}$),
$p_s\equiv(E_s=\sqrt{\vec{p}_s^2+m_p^2},\vec{p}_s)$ for
the spectator proton and
$p_x\equiv(E_X,\vec{p}_X)=(\nu+M_D-E_s,\vec{q}-\vec{p}_s)$ the center of mass
momentum
of the undetected produced hadronic system $X$.  The initial momentum of
the bound off-shell neutron is defined as $p_i\equiv p_D-p_s$.  After applying
all the abovementioned approximations and averaging over $\phi$ (the angle
between the electron and $\gamma^*,\vec{p}_s$ plane), we can write for the
differential cross section \cite{Cosyn:2010ux}:
\begin{multline}\label{eq:cross}
 \frac{d\sigma}{d\hat{x}dQ^2d^3p_s}=\frac{4\pi\alpha_{\text{EM}}}{Q^2\hat{x}}
\frac{|q|}{m_n}
\left(1-y-\frac { x^2y^2m_n^2} {Q^2}
\right)\left(\frac{Q^2}{|q|^2}+\frac{2\tan^2\
\frac{\theta_e}{2}}{1+R} \right)\left|\frac{\alpha_i}{
\alpha_q}+\frac{1}{2\hat{x}} \right|^{-1}\\\times\left[\left(\frac{\alpha_i}{
\alpha_q}+\frac{1}{2\hat{x}} \right)^2+\frac{p_T^2}{2Q^2}
\right]F_{2N}(\alpha_i,\hat{x},Q^2)S^D(p_s)\,.
\end{multline}
$\alpha_{\text{EM}}$ is the fine-structure constant,
$-Q^2=\nu^2-\vec{q}^2$ is the
four-momentum transfer,  $m_n$ the neutron mass, $y=\frac{\nu}{E_e}$, $\alpha_i
= \frac{2p_i^-}{M_D}$, $\alpha_q =
\frac{2q^-}{M_D}$,
$\hat{x}=\frac{Q^2}{2p_i\cdot q}$ is the Bjorken $x$ for a moving nucleon,
$\theta_e$ is the electron scattering angle, $R=\frac{\sigma_L}{\sigma_T}\approx
0.18$ is the ratio of the
longitudinal to transverse cross section for scattering off the nucleon,
and $F_{2N}$ is
the effective nucleon structure function, which is defined at $\hat{x}$ and
in principle could be modified due to the nuclear binding (see e.g. Ref.
\cite{Melnitchouk:1996vp}).  We use the SLAC parametrization of Ref.
\cite{PhysRevD.20.1471} for $F_{2N}$.

In Eq. (\ref{eq:cross}), the distorted spectral function $S^D(p_s)$ contains
the contributions from the plane-wave and FSI diagram and takes the following
form:
\begin{multline}
  S^D(p_s)= \frac{1}{3}\sum_{s_D,s_s,s_i} \left| \Psi^{s_D}(p_i
s_i, p_s
s_s) +\frac{i}{2}
\int\frac{d^2p_{s',\perp}}{(2\pi)^2}\frac{\beta(s,m_{X'})}{4\mid\vec{q}\mid\sqrt
{ E_sE_
{ s' } } }\right.\\\left.\times\left[ \langle X, p_s |
f^{\text{on}}_{X'N,XN}(s,t)| X', p_{s'}\rangle \Psi^{s_D}(\tilde{p}_{s'},
s_{i}, 
s_{s}) \right.\right.\\ \left.\left. -i \langle X, p_s |
f^{\text{off}}_{ X'N,XN}(s,t)| X' , p_{s'} \rangle \tilde{p}_{s',z}
\tilde{\Psi}^{s_D}(\tilde{p}_{s'},s_{i},
s_{s})\right]\right|^2\,,
\label{eq:distS}
\end{multline}
where $\Psi^{s_D}$ ($\tilde{\Psi}^{s_D}$) is the undistorted (distorted)
deuteron wave function with spin projection $s_D$, 
$\beta(s,m_{X'})=\sqrt{(s-(m_n-m_{X'} )^2)(s-(m_n+m_ { X' } )^2)}$ (with $s$
the Mandelstam variable of the rescattering process),  $m_{X'}$ is the
invariant mass of the produced hadronic state before the rescattering and
$\tilde{p}_{s',z}\equiv p_{s,z}-\Delta$, with
\begin{align} \label{eq:delta}
 \Delta&=\frac{\nu+M_D}{\mid\vec{q}\mid}(E_s-m_p)+\frac{m_X^2-\gamma}{
2\mid\vec { q } \mid } &\text{for}\, \gamma \leq m_X^2\,,\nonumber\\
\Delta&=\frac{\nu+M_D}{\mid\vec{q}\mid}(E_s-m_p) &\text{for}\, \gamma >
m_X^2\,,
\end{align}
where $\gamma= m_n^2+2m_n\nu-Q^2$ is the produced DIS
mass 
off the stationary nucleon.  The two regimes in Eq. (\ref{eq:delta}) originate
from the condition $m^2_{X}> m^2_{X^\prime}$, which can be inferred from the
approximate conservation law for  the ``$-$'' 
component in high energy small angle
scatterings \cite{Cosyn:2010ux}.

The on-shell rescattering amplitude takes the form of Eq. (\ref{eq:ampl}).  For
the off-shell amplitude
$f^{\text{off}}_{ X'N,XN}$ there is
no clear prescription, but  following our main goal of studying the
semi-inclusive DIS 
based only on basic properties of the high-energy scattering we identify two
extreme cases 
for off-shell part of the rescattering amplitude, one when it is taken to be
zero 
({\em no off-shell FSI}) and the other in which off-shell amplitude is assumed
to be equal to 
the on-shell amplitude $f^{\text{on}}_{X'N,XN}$ referred as {\em maximal
off-shell FSI}.  A third approach is to parametrize the amplitude as
\begin{equation}\label{eq:offrescatter}
 f^{\text{off}}_{ X'N,XN}=f^{\text{on}}_{ X'N,XN} e^{-\mu(x,Q^2)t},
\end{equation}
with $\mu$ an extra parameter that can be fitted, providing a measure of the
suppression of the off-shell amplitude.  This approach is referred to as
\emph{fitted off-shell FSI}.  

\section{Results}
\begin{figure}[!htb]
  \includegraphics[width=0.98\textwidth,height=0.8\textheight]{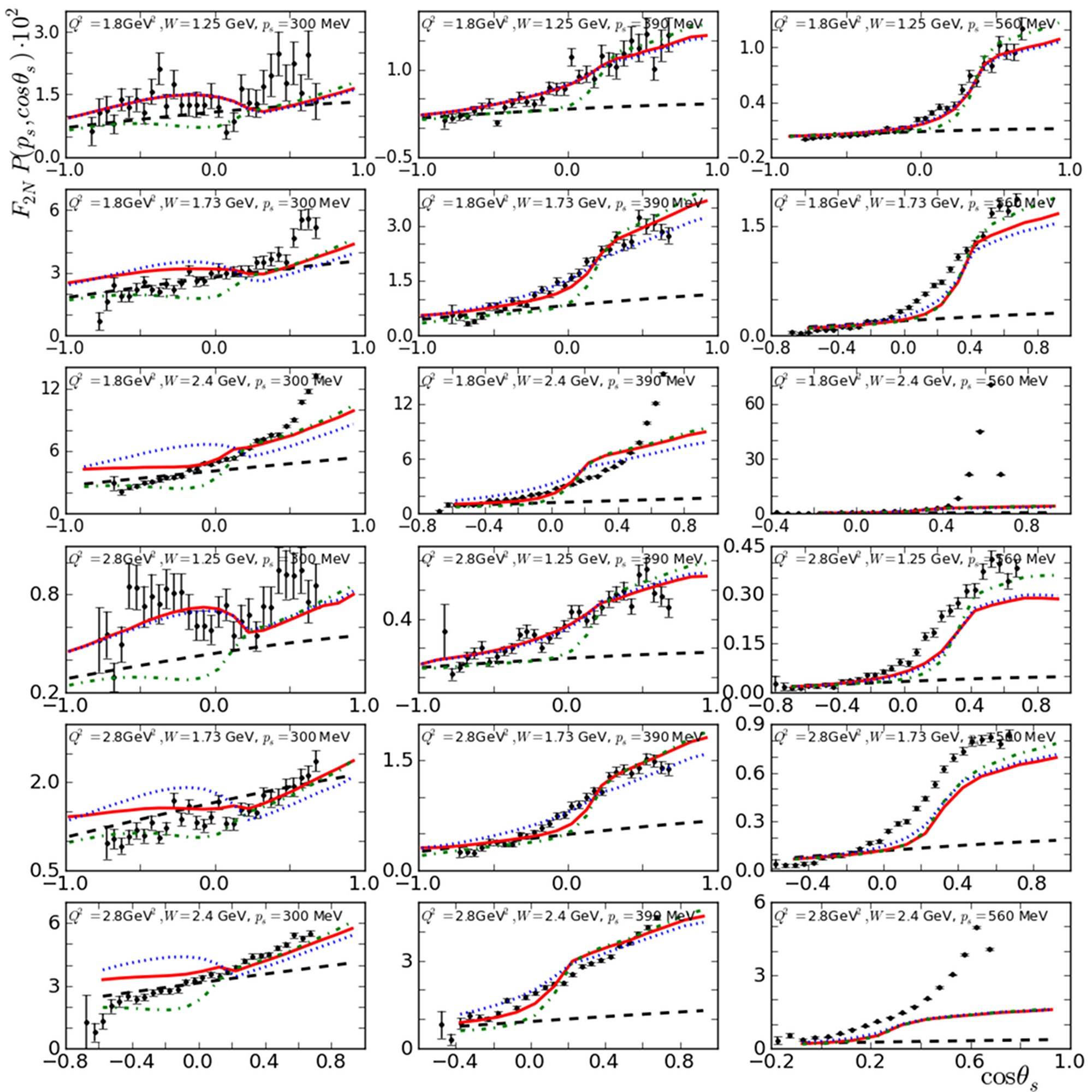}
\caption{(Color online) Comparison between the Deeps data
\cite{Klimenko:2005zz} and model calculations.  The
dashed black curve is a
plane-wave calculation, the other include final-state interactions.  The
effective total cross section and slope parameter in the final-state
interaction amplitude are fitted parameters for each $Q^2,W$, the
real part is fixed at $\epsilon=-0.5$.  The dot-dashed green curve only
considers
on-shell
rescattering, the dotted blue curve has an off-shell rescattering amplitude
equal to
the on-shell one and the full red curve uses the off-shell
parameterization of Eq.
(\ref{eq:offrescatter})}
\label{fig:fit}       
\end{figure}

Fig. \ref{fig:fit} compares our model calculations with a selection of data
taken in the Deeps experiment (for a definition of the spectral function
$P(p_s,\cos\theta_s)$ see Ref. \cite{Klimenko:2005zz}).  For each $Q^2,W$, the
total cross section $\sigma$ and slope parameter $B$ in Eq. (\ref{eq:cross})
were taken as free parameters and fitted to the data for all $p_s$ values.  In
the case of the \emph{fitted off-shell FSI}, $\mu$ was taken as an extra free
parameter.   As can be observed in the figures, the calculations including
FSI do a fair job of describing the data over the kinematic range of the
experiment.  The differences between the three different off-shell descriptions
becomes smaller with higher $p_s$. At $p_s=300$ MeV the plane-wave and FSI
amplitudes are of equal magnitude, making the final result very sensitive to the
different descriptions, especially noticeable in the backward region. At
this $p_s$ value, there is
also an oscillating structure in the data which disappears with higher $W$, but
is still present in the model results.

At the higher spectator momenta, the calculations including FSI continously
grow for angles in the forward direction, clearly different from the fairly
flat behavior of the plane-wave calculations.  This high FSI contribution in
the forward region is a consequence of the structure of the phase factor in Eq.
(\ref{eq:delta}).  At the highest $p_s$, the calculations systematically
underestimate the data, hinting at a breakdown of the factorization used in the
derivation of Eq. (\ref{eq:cross}).  Over the whole kinematic range of the
data, the \emph{fitted off-shell FSI} calculations are more in agreement with
the \emph{no off-shell} than the \emph{maximal off-shell} ones, pointing in the
direction of a largely suppressed off-shell rescattering amplitude.

\begin{figure}[!htb]
  \includegraphics[width=0.7\textwidth]{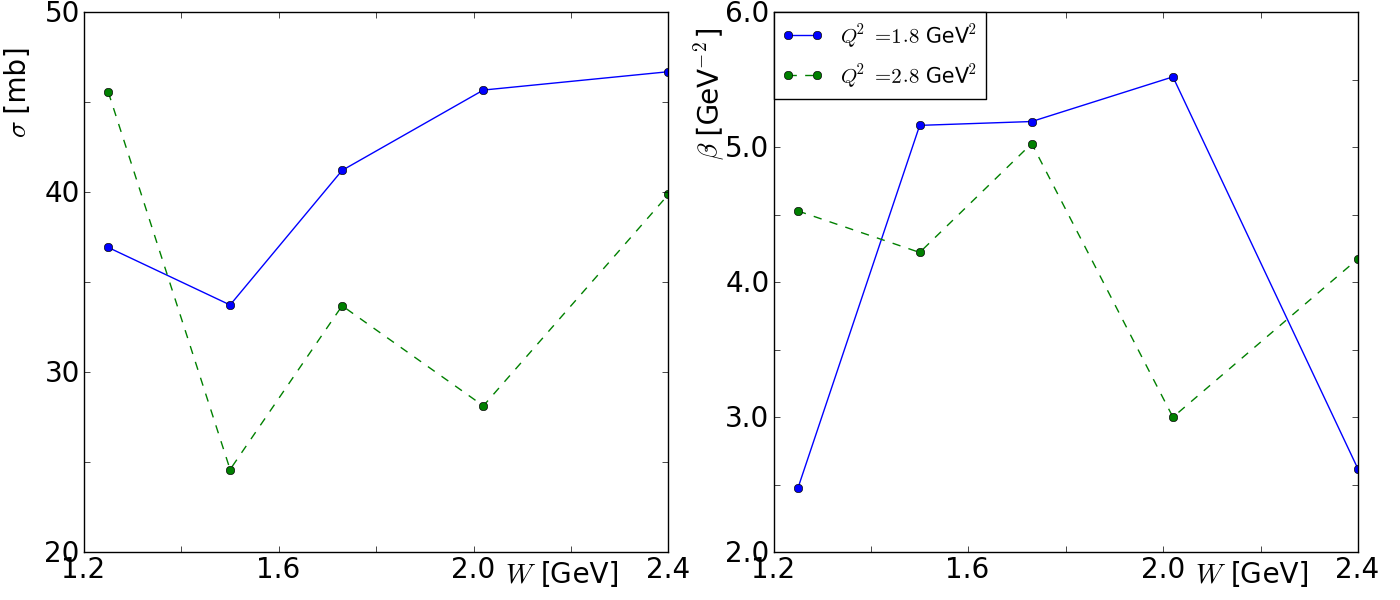}
\caption{(Color online) The fitted values of effective cross section $\sigma$
and slope factor $B$ for the \emph{no off-shell FSI} calculations
used in Fig. \ref{fig:fit} as a function of the
invariant mass $W$.  Full blue curve is for $Q^2=1.8~\text{GeV}^2$,
the dashed green curve for $Q^2=2.8~\text{GeV}^2$.}
\label{fig:fit2paramon}       
\end{figure}

\begin{figure}[!htb]
  \includegraphics[width=0.7\textwidth]{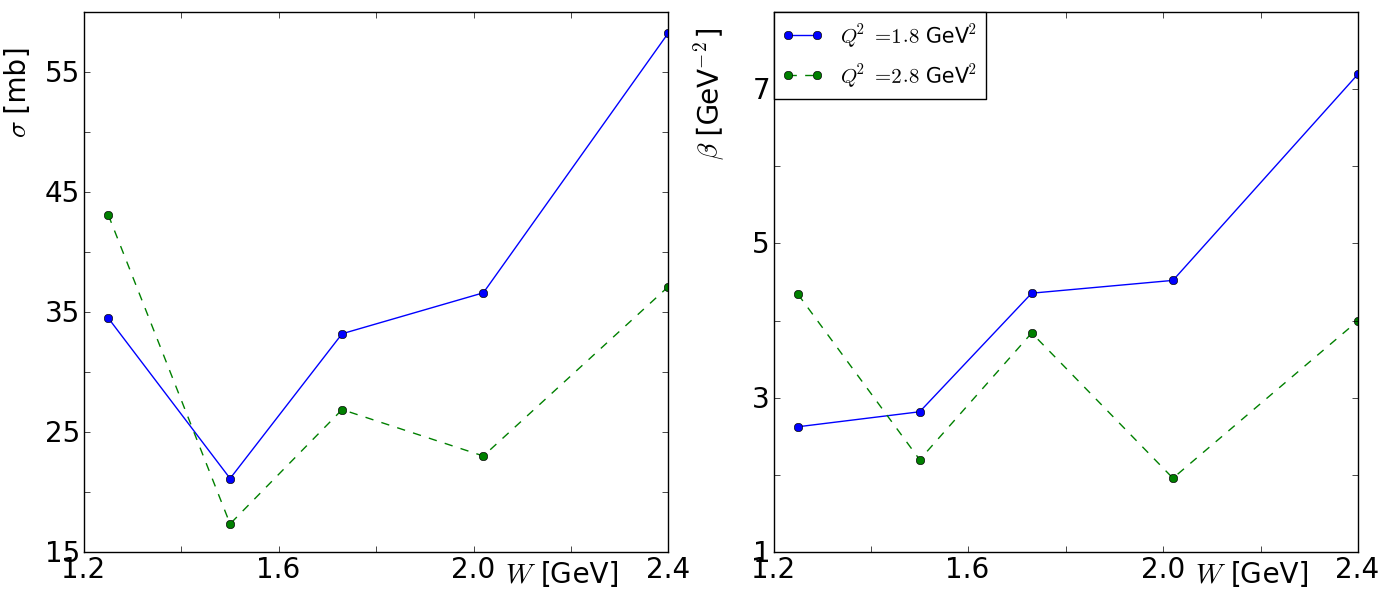}
\caption{(Color online) The fitted values of effective cross section $\sigma$
and slope factor $B$ for the \emph{maximal off-shell} FSI calculations
used in Fig. \ref{fig:fit} as a function of the
invariant mass $W$.  Full blue curve is for $Q^2=1.8~\text{GeV}^2$,
the dashed green curve for $Q^2=2.8~\text{GeV}^2$.}
\label{fig:fit2param}       
\end{figure}

The values of the fitted total cross section $\sigma$ and slope parameter $B$
used in Fig. \ref{fig:fit} are shown in Figs. \ref{fig:fit2paramon} and
\ref{fig:fit2param} for the \emph{no off-shell FSI} and \emph{maximal off-shell
FSI} calculations.  After an initial peak at $W=1.2$ GeV, corresponding with
the creation of a $\Delta$, the value of $\sigma$ drops to around 25 mb and
rises with increasing $W$.  This agrees with the picture of the
increased creation of hadronic constituents as $W$ increases.  With increasing
$Q^2$, the
value of the $(XN)$ cross section parameter also becomes consistently smaller in
this region, indicating reduced final-state interactions.  This could point at
the onset of a color transparency signal, but more data points at other $Q^2$
and $W$ values are needed to make more substantial claims.  The value of the
slope parameter $B$ is also largely correlated with the
$Q^2,W$-dependence  of the $\sigma$ parameter.

\section{Neutron structure at high $x$}

We now use the Deeps data and our model calculations to outline a method to
extract the neutron structure function $F_2$ at high $x$ from kinematics at low
$x$.  It is based on an analytical continuation of the amplitude to the
unphysical limit $t'=m_n^2 - p_i^2 \rightarrow 0$, which corresponds to an
on-shell neutron.  Due to the low binding energy of the deuteron this
singularity of the amplitude is very close to the physical region.  It is
analogous to the Chow-Low procedure applied in the extraction of $\pi\pi$ and
$NN$ cross sections from $p(\pi,p)X$ and $d(n,n)pn$ reactions
\cite{Chew:1958wd}.  It can be shown that the plane-wave part of the distorted
spectral function in Eq. (\ref{eq:distS}) has a double pole in $t'$, while the
FSI part hasn't \cite{Sargsian:2005rm}.  This allows us to extract the $F_2$
structure function as
\begin{equation}
 \displaystyle\lim_{t'\to
0}F_{2N}^{\text{extr}}(Q^2,\hat{x},t')=\frac{t'^2}{[\text{Res}(\Psi_D(t'=0))]^2}
\frac{F_{L}^{D,\text{exp}}(x,Q^2)+v_TF_
{T}^{D,\text{exp}}(x,Q^2)}{\frac{2\hat{x}\nu}{m_n} \left [
\left(\frac{\alpha_i}{
\alpha_q}+\frac{1}{2\hat{x}} \right)^2+\frac{p_T^2}{2Q^2}\left(\frac{Q^2}{\mid
q\mid^2}+\frac{2\tan^2\
\frac{\theta_e}{2}}{1+R} \right)
\right]}\,,\label{eq:extrapolate}
\end{equation}
where $F_{L,T}^{D,\text{exp}}$ are the experimentally measured deuteron
structure functions.  The rhs of Eq. (\ref{eq:extrapolate}) has a
quadratic dependence on $t'$.

\begin{figure}[!htb]
  \includegraphics[width=0.7\textwidth]{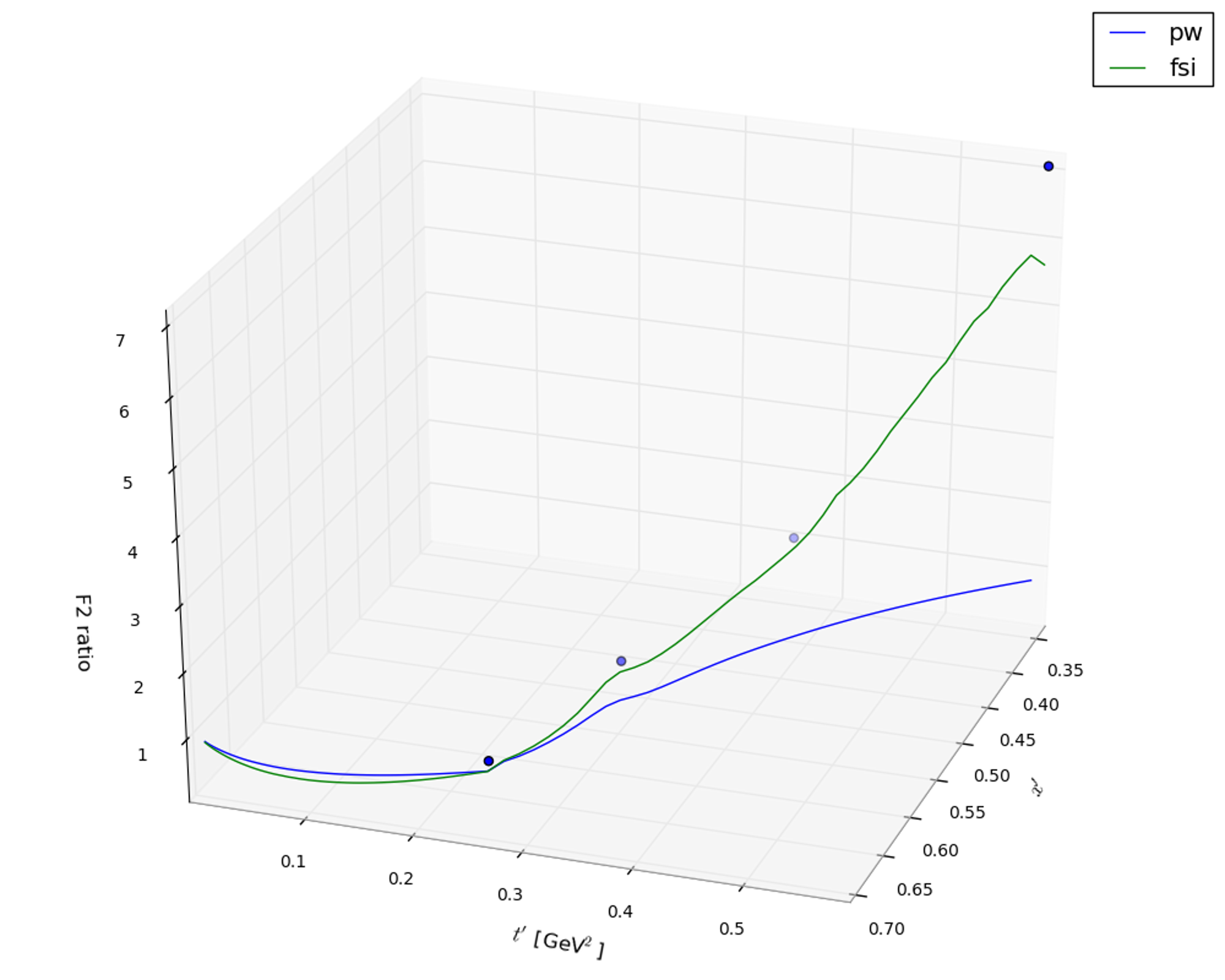}
\caption{(Color online) Example of a trajectory in $t',\hat{x}$ used in the
extrapolation procedure for the neutron structure function $F_2$.  The $z$-axis
shows the ratio of Eq. (\ref{eq:extrapolate}) to the parametrization of Ref.
\cite{PhysRevD.20.1471}.  The blue (green) curve shows the plane-wave (FSI)
model calculations along the trajectory, blue dots are JLab Deeps data
\cite{Klimenko:2005zz}.}
\label{fig:traj}       
\end{figure}

The approach we now take is to take a trajectory through the Deeps data at a
fixed angle that starts at high $W, p_s$ and goes to low $W, p_s$
values.  This translates into a descending series in $t'\rightarrow 0$  and an
ascending series in $\hat{x} \rightarrow x$, as is shown in Fig.
\ref{fig:traj}.  In this manner, we can use measurements at low values of
$\hat{x}$ - where $F_{2N}$ is well known - to extrapolate to regions with high
$x$.  

\begin{figure}[!htb]
  \includegraphics[width=0.6\textwidth]{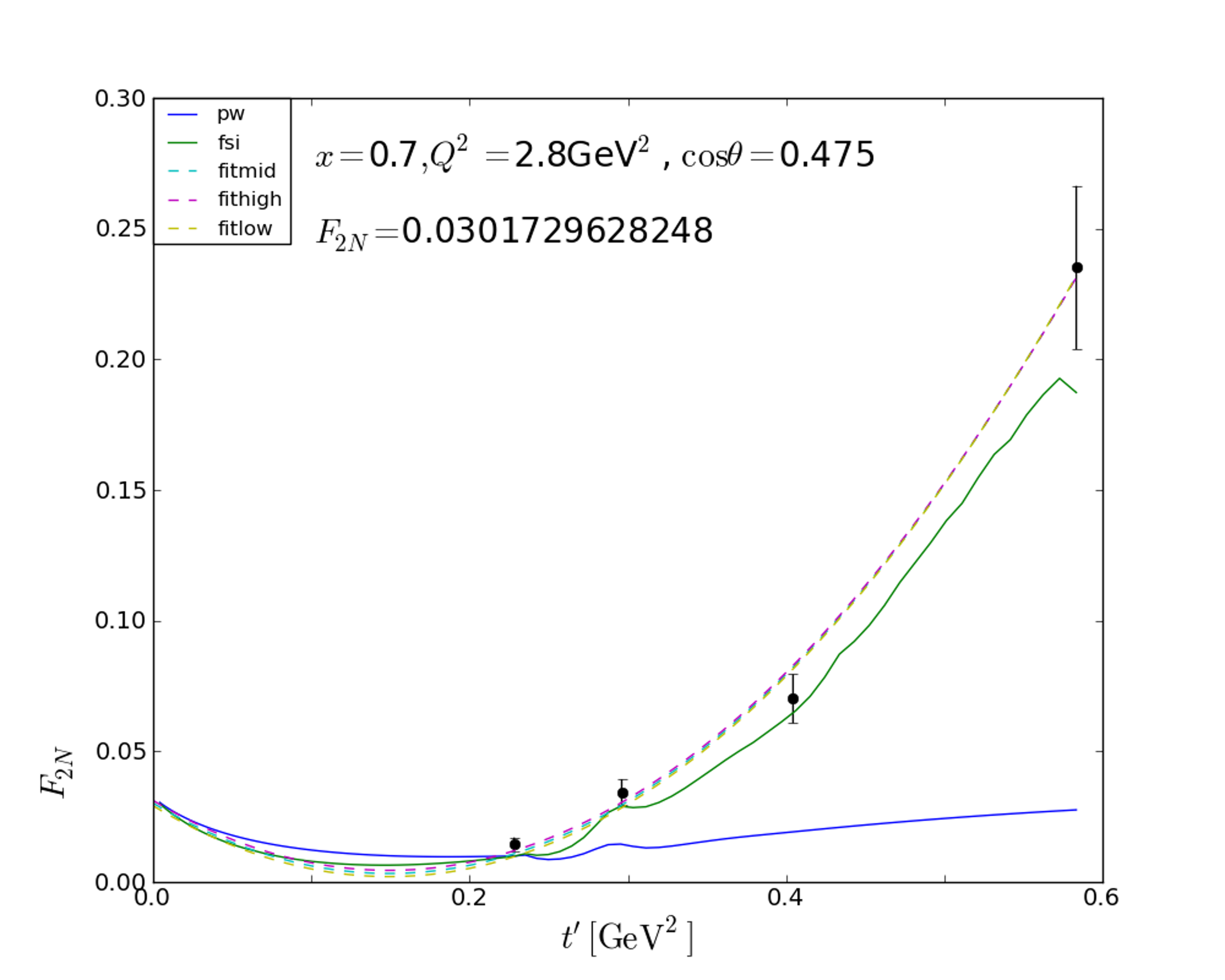}
\caption{(Color online) Example of a extrapolation for $F_{2N}$ at $x=0.7$.  The
blue (green) curve shows the plane-wave (FSI)
model calculations along the trajectory, black dots are JLab Deeps data
\cite{Klimenko:2005zz}.  The dashed curves are a quadratic fit to the data
points, the intercept at $t'=0$ yields the $F_{2N}(x=0.7)$ value.}
\label{fig:extra}       
\end{figure}

Fig. \ref{fig:extra} shows an example of a quadratic fit to the Deeps data,
extrapolated to $t'=0$ to obtain the neutron structure function $F_{2N}$ at
Bjorken $x=0.7$.  As the available data sits on one arm of the quadratic curve,
it's not straightforward to do the extrapolation.    Here, we imposed an
extra constraint on the quadratic fit, that the $t'$ value of its minimum has to
coincide with the minimum of the FSI curve along the trajectory.  The position
of this minimum is related to the size of the FSI term in the distorted
spectral function which the model does a good of describing for the kinematics
along this trajectory.  The three different fit curves were obtained by
taking the data point and error bar values respectively for the lowest
$t'$ data point.  With the current data set, the method doesn't provide
a robust prediction for the high $x$ neutron structure function, as
trajectories with different spectator angles yield a range of values for
$F_{2N}$.  Ideally, more data at
lower $p_s$ values would provide better constraints and yield better
predictions.

\begin{theacknowledgments}
This work is supported by the Research Foundation Flanders as well as by the  
U.S. Department of Energy Grant 
under Contract DE-FG02-01ER41172.
\end{theacknowledgments}



\bibliographystyle{aipproc}   

\begin{thebibliography}{22}
\expandafter\ifx\csname natexlab\endcsname\relax\def\natexlab#1{#1}\fi
\providecommand{\enquote}[1]{``#1''}
\expandafter\ifx\csname url\endcsname\relax
  \def\url#1{\texttt{#1}}\fi
\expandafter\ifx\csname urlprefix\endcsname\relax\def\urlprefix{URL }\fi
\providecommand{\eprint}[2][]{\url{#2}}

\bibitem[Klimenko et~al.(2006)]{Klimenko:2005zz}
A.~V. Klimenko, et~al., \emph{Phys. Rev.} \textbf{C73}, 035212 (2006),
  \eprint{nucl-ex/0510032}.

\bibitem[Fenker et~al.(2003)]{Bonus:2003}
H.~Fenker, C.~Keppel, S.~Kuhn, and W.~Melnitchouk  (2003),
  \urlprefix\url{http://jlab.org/exp_prog/CEBAF_EXP/E03012.html},
  jLAB-PR-03-012.

\bibitem[Simula(1996)]{Simula:1996xk}
S.~Simula, \emph{Phys. Lett.} \textbf{B387}, 245--252 (1996),
  \eprint{nucl-th/9605024}.

\bibitem[Melnitchouk et~al.(1997)]{Melnitchouk:1996vp}
W.~Melnitchouk, M.~Sargsian, and M.~I. Strikman, \emph{Z. Phys.} \textbf{A359},
  99--109 (1997), \eprint{nucl-th/9609048}.

\bibitem[Sargsian and Strikman(2006)]{Sargsian:2005rm}
M.~Sargsian, and M.~Strikman, \emph{Phys. Lett.} \textbf{B639}, 223--231
  (2006), \eprint{hep-ph/0511054}.

\bibitem[Ciofi~degli Atti et~al.(1999)]{CiofidegliAtti:1999kp}
C.~Ciofi~degli Atti, L.~P. Kaptari, and S.~Scopetta, \emph{Eur. Phys. J.}
  \textbf{A5}, 191--207 (1999), \eprint{hep-ph/9904486}.

\bibitem[Ciofi~degli Atti and Kopeliovich(2003)]{CiofidegliAtti:2002as}
C.~Ciofi~degli Atti, and B.~Z. Kopeliovich, \emph{Eur. Phys. J.} \textbf{A17},
  133--144 (2003), \eprint{nucl-th/0207001}.

\bibitem[Ciofi~degli Atti et~al.(2004)]{CiofidegliAtti:2003pb}
C.~Ciofi~degli Atti, L.~P. Kaptari, and B.~Z. Kopeliovich, \emph{Eur. Phys. J.}
  \textbf{A19}, 145--151 (2004), \eprint{nucl-th/0307052}.

\bibitem[Palli et~al.(2009)]{Palli:2009it}
V.~Palli, C.~Ciofi~degli Atti, L.~P. Kaptari, C.~B. Mezzetti, and M.~Alvioli,
  \emph{Phys. Rev.} \textbf{C80}, 054610 (2009), \eprint{0911.1377}.

\bibitem[Atti and Kaptari(2010)]{Atti:2010yf}
C.~C.~d. Atti, and L.~P. Kaptari  (2010), \eprint{1011.5960}.

\bibitem[Cosyn and Sargsian(2010)]{Cosyn:2010ux}
W.~Cosyn, and M.~Sargsian  (2010), \eprint{1012.0293}.

\bibitem[Sargsian et~al.(2002)]{Sargsian:2001gu}
M.~M. Sargsian, S.~Simula, and M.~I. Strikman, \emph{Phys. Rev.} \textbf{C66},
  024001 (2002), \eprint{nucl-th/0105052}.

\bibitem[Sargsian(2010)]{Sargsian:2009hf}
M.~M. Sargsian, \emph{Phys. Rev.} \textbf{C82}, 014612 (2010),
  \eprint{0910.2016}.

\bibitem[Sargsian(2001)]{Sargsian:2001ax}
M.~M. Sargsian, \emph{Int. J. Mod. Phys.} \textbf{E10}, 405--458 (2001),
  \eprint{nucl-th/0110053}.

\bibitem[Sargsian et~al.(2003)]{Sargsian:2002wc}
M.~M. Sargsian, et~al., \emph{J. Phys.} \textbf{G29}, R1 (2003),
  \eprint{nucl-th/0210025}.

\bibitem[Frankfurt and Strikman(1981)]{Frankfurt1981215}
L.~L. Frankfurt, and M.~I. Strikman, \emph{Physics Reports} \textbf{76}, 215 --
  347 (1981).

\bibitem[Frankfurt and Strikman(1976)]{Frankfurt:1976gz}
L.~L. Frankfurt, and M.~I. Strikman, \emph{Phys. Lett.} \textbf{B64}, 433--434
  (1976).

\bibitem[Frankfurt and Strikman(1987)]{Frankfurt1987254}
L.~L. Frankfurt, and M.~I. Strikman, \emph{Physics Letters B} \textbf{183}, 254
  -- 258 (1987).

\bibitem[Landshoff and Polkinghorne(1978)]{Landshoff:1977pg}
P.~V. Landshoff, and J.~C. Polkinghorne, \emph{Phys. Rev.} \textbf{D18}, 153
  (1978).

\bibitem[Jeschonnek(2001)]{Jeschonnek:2000nh}
S.~Jeschonnek, \emph{Phys. Rev.} \textbf{C63}, 034609 (2001),
  \eprint{nucl-th/0009086}.

\bibitem[Bodek et~al.(1979)]{PhysRevD.20.1471}
A.~Bodek, M.~Breidenbach, D.~L. Dubin, J.~E. Elias, J.~I. Friedman, H.~W.
  Kendall, J.~S. Poucher, E.~M. Riordan, M.~R. Sogard, D.~H. Coward, and D.~J.
  Sherden, \emph{Phys. Rev. D} \textbf{20}, 1471--1552 (1979).

\bibitem[Chew and Low(1959)]{Chew:1958wd}
G.~F. Chew, and F.~E. Low, \emph{Phys. Rev.} \textbf{113}, 1640--1648 (1959).

\end{thebibliography}


\end{document}